\documentclass[twocolumn,english,showpacs,preprintnumbers,amsmath,amssymb,floatfix]{revtex4}
\usepackage[T1]{fontenc}
\usepackage[latin9]{inputenc}
\usepackage{color}
\usepackage{array}
\usepackage{amstext}
\usepackage{graphicx}
\usepackage{esint}

\makeatletter

\@ifundefined{textcolor}{}
{%
 \definecolor{BLACK}{gray}{0}
 \definecolor{WHITE}{gray}{1}
 \definecolor{RED}{rgb}{1,0,0}
 \definecolor{GREEN}{rgb}{0,1,0}
 \definecolor{BLUE}{rgb}{0,0,1}
 \definecolor{CYAN}{cmyk}{1,0,0,0}
 \definecolor{MAGENTA}{cmyk}{0,1,0,0}
 \definecolor{YELLOW}{cmyk}{0,0,1,0}
 }

\@ifundefined{definecolor}
 {\usepackage{color}}{}
\@ifundefined{definecolor}
 {\usepackage{color}}{}
\makeatother

\makeatother

\usepackage{babel}

\begin{document}

\title{B-mesons from top-quark decay in presence of the charged-Higgs\\
 boson in the Zero-Mass Variable-Flavor-Number Scheme}

\author{S. Mohammad Moosavi Nejad $^{a,b}$}

\email{mmoosavi@yazduni.ac.ir}

\affiliation{$^{(a)}$Faculty of Physics, Yazd University, P.O. Box
89195-741, Yazd, Iran}

\affiliation{$^{(b)}$School of Particles and Accelerators,
Institute for Research in Fundamental Sciences (IPM), P.O.Box
19395-5531, Tehran, Iran}

\date{\today}

\begin{abstract}

We study the energy spectrum of the inclusive bottom-flavored mesons in top-quark decay into a charged-Higgs-boson and a massless bottom quark  at next-to-leading order (NLO) in the two-Higgs-doublet model. To extract the result we work in the Zero-Mass Variable-Flavor-Number scheme(ZM-VFNs) using realistic nonperturbative fragmentation functions obtained through a global fit to $e^+e^-$ data from  CERN LEP1 and SLAC SLC  on the Z-boson resonance.  We study both the contribution of the bottom-quark fragmentation and the  gluon fragmentation  to produce the  bottom-flavored meson (B-Meson). We find that the contribution of the gluon leads to an appreciable reduction in  decay rate  at low values of the B-meson energy. It means the NLO corrections are significant.

\end{abstract}

%\pacs{Valid PACS appear here}% PACS, the Physics and Astronomy
                             % Classification Scheme.
%\keywords{Suggested keywords}%Use showkeys class option if keyword
                              %display desired
\maketitle

\section{Introduction}

Top quark  is the heaviest elementary particle with a mass of $172.0$ GeV \cite{Nakamura:2010zzi}. Top's large mass is a reason to rapid decay  so that it has no time to
hadronize and if it were not for the confinement of color, the top quark could be considered as a free particle. This property allows the top quark  to behave like a real particle and one can safely describe its decay in perturbative theory. The Large Hadron Collider (LHC) is a superlative top
factory with 90 million $t\bar t$-pairs per year which will allow us to perform accurate studies of the top-quark properties, such as its mass $m_t$, total decay width $\Gamma_t$, the effective weak mixing angle, and elements
$V_{tq}$ of the Cabibbo-Kobayashi-Maskawa (CKM) \cite{Cabibbo:1963yz} quark mixing matrix. The theoretical aspects of top-quark physics at the LHC are summarized
in a recent paper \cite{Bernreuther:2008ju}.

Since $|V_{tb}|\approx 1$,  top quarks  almost exclusively decay to bottom quarks, via $t \rightarrow bW^+$ within the Standard Model (SM) theory and in beyond-the-SM theories  with an extended Higgs sector, top quarks decay via $t \rightarrow bH^+$.\\
 Many extensions of the Standard Model contain more than one Higgs doublet, and the new degrees of freedom appear as extra Higgs scalars. For example, in the supersymmetric SM, at least  two Higgs doublets are needed so as to cancel gauge anomalies and to generate masses for both up- and down-type quarks \cite{Gunion}; in the Weinberg model of CP violation, at least three Higgs doublets are needed in order to have spontaneous CP violation(see \cite{Li} and references therein).  Both neutral and charged physical Higgs bosons exist in all these extensions of the SM. The observation of charged Higgs bosons, $H^\pm$, would indicate physics beyond the SM. For the sake of simplicity, we will limit ourselves throughout this paper to the decay $t \rightarrow bH^+$ in a model with two-Higgs-doublet, in which case there is only one physical charged Higgs boson remaining after spontaneous symmetry breaking.  New results of a search for the charged Higgs bosons in proton-proton collision at a center-of-mass energy of $\sqrt{s}=7$ TeV are reported by the ATLAS Collaboration \cite{ATLAS} where the $\tau+$jets channel in $t\bar t$ decays is used with a hadronically decaying $\tau$ lepton in the final state. The reported data leads to a limit on the product of branching ratios $BR(t \rightarrow bH^\pm)\times BR(H^\pm \rightarrow \tau\nu)$ of 0.03-0.10 for $H^\pm$ masses in the range $90$ GeV$<m_{H^\pm}<160$ GeV.\\
  As it is mentioned in Ref.~\cite{Ali:2011qf}, a clear separation between the $t \rightarrow bW^+$ and $t \rightarrow bH^+$ can be achieved in both the $t\bar{t}X$ pair production and the  $t/\bar{t}X$ single top production at the LHC. The QCD corrections to the decay rate $t\rightarrow b+W^+$ are known at  next-to-next-to-leading order \cite{Czarnecki:1998qc} and the NLO electroweak corrections were found in Ref.~\cite{Denner:1990ns}. In this paper, we evaluate the first order QCD corrections to the decay of an unpolarized top quark into a charged Higgs boson.

Since  bottom quarks hadronize, via $b\to B+X$, therefore  the decay process $t\to B H^++X$ is of prime importance and the particular purpose of this paper is to make predictions for the energy spectrum of b-flavored mesons in top decay. This measurement will be important for future tests of the Higgs coupling in the minimal supersymmetric standard model (MSSM) at LHC. The hadronization of the bottom quark was considered in the NLO QCD analyses of  the decay $t\rightarrow bW^+$ in Refs.~\cite{Corcella:1,Corcella:2}.  As it is shown in \cite{do}, the hadronization of the bottom quark is identified to be the largest source of uncertainty in  measurement of the top-quark mass.  \\
To  study the distributions in the $B$-meson scaled-energy
$x_B$, we employ  the massless scheme or  zero-mass variable-flavor-number (ZM-VFN) scheme \cite{jm} in the top-quark rest frame.  In this scheme, the zero mass parton approximation is also applied  to the bottom quark and the non-zero value of the b-quark mass only enter through the initial condition of the nonperturbative fragmentation function(non-FF). Nonperturbative FF is describing the hadronization process $b\rightarrow B+X$ and is subject to Dokshitzer-Gribov-Lipatov-Alteralli-Parisi(DGLAP) \cite{dglap} evolution and it is scale dependent.

This paper is organized as follows.
In Sec.~\ref{sec:one}, we study the inclusive production of a meson from top-quark decay considering the factorization theorem and DGLAP equations.
In Sec.~\ref{sec:two}, we present the calculation of the ${\cal O}(\alpha_S)$ QCD corrections to the tree-level rate of $t\rightarrow bH^+$. We work in ZM-VFN scheme neglecting the b-quark mass in our QCD corrections but will retain the arbitrary value of $m_{H^+}$.
In Sec.~\ref{sec:four}, we present our numerical analysis.
In Sec.~\ref{sec:five},  our conclusions are summarized.

\boldmath
\section{Formalism}
\label{sec:one}
\unboldmath

We study the inclusive production of a B-meson from top-quark decay
\begin{eqnarray}\label{pros}
t\rightarrow b+H^+ (g)\rightarrow BH^++X,
\end{eqnarray}
providing that the top-quark mass $m_t$, bottom-quark mass $m_b$ and the charged Higgs boson mass $m_H^+$ satisfy $m_t>m_b+m_H^+$ . The gluon in Eq.~(\ref{pros}) contributes to the real radiation at NLO and both the $b$ quark and the gluon may hadronize to the $B$ meson. In the equation above,  $X$ stands for the unobserved final state.

 If we denote the four-momenta of top-quark, b quark, gluon and B meson by $p_t, p_b, p_g$ and $p_B$, respectively, therefore  in the top-quark rest frame the $b$ quark, gluon, and $B$ meson have energies $E_i=p_t\cdot p_i/m_t  (i=b,g,B)$, where $m_B\le E_B\le (m_t^2+m_B^2-m^2_{H^+})/(2m_t) $ and  $0\le (E_b,E_g)\le (m_t^2-m^2_{H^+})/(2m_t) $.
It is convenient to introduce the scaled energy fractions  $x_i=E_i/E_b^\text{max}$ ($i=b,g,B$).\\
We wish to calculate the partial decay width of process~(\ref{pros}) differential in $x_B$, $d\Gamma/dx_B$, at NLO in the ZM-VFN scheme. According to the factorization  theorem of the QCD-improved parton model \cite{jc}, the energy distribution of a hadron B can be expressed as the convolution of the parton-level spectrum with the nonperturbative fragmentation function $D_a(z,\mu_F)$,  describing the hadronization $a\rightarrow B$,
\begin{equation}
\frac{d\Gamma}{dx_B}=\sum_{a=b,g}\int_{x_a^\text{min}}^{x_a^\text{max}}
\frac{dx_a}{x_a}\,\frac{d\hat\Gamma_a}{dx_a}(\mu_R,\mu_F)
D_a\left(\frac{x_B}{x_a},\mu_F\right),
\label{eq:master}
\end{equation}
where $d\hat\Gamma_a/dx_a$ is the parton-level differential width  of the process $t\to a+X$, with $X$ comprising the $H^+$ boson and any other parton. Here, $\mu_F$ and $\mu_R$ are the factorization and the renormalization scales, respectively. At NLO, the scale $\mu_R$ is associated with the renormalization of the strong coupling constant. In principle, one can use two different values for the factorization and renormalization scales; however, a choice often made consists of setting $\mu_R=\mu_F$ and we shall adopt this convention for most of the result which we shall show.\\
 In next section, we present our analytic results for $d\hat\Gamma_a/dx_a (a=b,g)$ at NLO in ZM-VFNS.

\boldmath
\section{Analytic results for $d\hat\Gamma_a/dx_a$}
\label{sec:two}
\unboldmath

\boldmath
\subsection{Tree-Level Rate of $t\rightarrow bH^+$ in ZM-VFNS}
\unboldmath
The coupling of the charged Higgs boson to the top and bottom quark in the minimal supersymmetric standard model(MSSM) can either be expressed as a superposition of  scalar and pseudoscaler coupling factors or as a superposition of right- and left-chiral coupling factors \cite{higg}. Adopting the first approach, the Born term amplitude for the process $t\rightarrow b+H^+$ can be parametrized as \textit{$M_0=\bar{u_b}(a+b\gamma_5)u_t$}, and the second scheme leads to the Born amplitude \textit{$M_0=\bar{u_b}\{g_t(1+\gamma_5)/2+g_b(1-\gamma_5)/2\}u_t$} where $a=(g_t+g_b)/2 $ and $b=(g_t-g_b)/2$.
In a model with two Higgs doublets and generic coupling to all the quarks, it is difficult to avoid flavor-changing neutral currents. We, therefore, limit ourselves to models that naturally stop these problems by restricting the Higgs coupling. As it is explained in Ref.~\cite{higg}, the first possibility  is to have  the doublet $H_1$ coupling  to all bosons and the doublet $H_2$ coupling to all the quarks (model $I$). This leads to the coupling factors
\begin{eqnarray}
a&=&\frac{g_\omega}{2\sqrt{2}m_W}V_{tb}(m_t-m_b)\cot\beta,\nonumber\\
b&=&\frac{g_\omega}{2\sqrt{2}m_W}V_{tb}(m_t+m_b)\cot\beta.
\end{eqnarray}
The second possibility  is to have the  $H_2$ couple to the right-chiral up-type quarks ($u_R, c_R, t_R$), and the $H_1$ couple to  the right-chiral down-type quarks (model $II$). This model leads to the coupling factors
\begin{eqnarray}
a&=&\frac{g_\omega}{2\sqrt{2}m_W}V_{tb}(m_t \cot\beta+m_b\tan\beta),\nonumber\\
b&=&\frac{g_\omega}{2\sqrt{2}m_W}V_{tb}(m_t \cot\beta-m_b\tan\beta).
\end{eqnarray}
In equations above, $\tan\beta=\nu_2/\nu_1$ is  the ratio of the vacuum expectation values of the two electrically neutral components of the two Higgs doublets and the weak coupling factor $g_w$ is related to the Fermi's constant $G_F$ by $g_\omega^2=4\sqrt{2}m_W^2G_F$. \\
The total decay width of $t\rightarrow H^+b$ at LO is given by
\begin{eqnarray}
\Gamma_0&=&\frac{m_t(a^2+b^2)}{16\pi }(1+\frac{m_b^2}{m_t^2}-\frac{m_{H^+}^2}{m_t^2})\times\nonumber\\
&&\lambda^{\frac{1}{2}}(1,\frac{m_b^2}{m_t^2},\frac{m_{H^+}^2}{m_t^2})\bigg\{1+\frac{2m_bm_t}{m_t^2+m_b^2-m_{H^+}^2}\frac{a^2-b^2}{a^2+b^2}\bigg\},\nonumber
\end{eqnarray}
where $\lambda(a,b,c)=a^2+b^2+c^2-2(ab+bc+ca)$ is the  K\"all\'en function.
In the limit of vanishing b-quark mass, $a=b$ in model $I$, therefore the tree level  decay width simplifies to
\begin{eqnarray}\label{gamma}
\hat\Gamma_0=\frac{m_t^3}{8\sqrt{2}\pi}G_F|V_{tb}|^2(1-\frac{m_{H^+}^2}{m_t^2})^2\cot^2\beta.
\end{eqnarray}
For model $II$ one has
\begin{eqnarray}
\frac{a^2-b^2}{a^2+b^2}=2\frac{m_bm_t}{m_t^2\cot^2\beta+m_b^2\tan^2\beta},\nonumber\\
 \end{eqnarray}
 when the left-chiral coupling term, proportional to $m_b\tan\beta$, can become comparable to the right-chiral coupling term  $m_t\cot\beta$, one cannot therefore naively set $m_b=0$ in all expressions. For example, if we take  $m_b=4.90$ GeV, $m_t=172.0$ GeV, $m_{H^+}=120$ GeV and $\tan\beta\approx 10$ thus the second term in the curly brackets can become as large as ${\cal O}(6\%)$ in model $II$. In this paper we adopt, with good approximation,  the Born term presented in Eq.~(\ref{gamma}) in both models when  $m_b\rightarrow 0$,   more detail can be found in  Ref.~\cite{kadeer}.

In the following,  we discuss the calculation of the ${\cal O}(\alpha_S)$ QCD corrections to the tree-level decay rate of $t\rightarrow b+H^+$ and we present the parton-level expressions for $d\Gamma(t\rightarrow BH^++X)/dx_B$ at NLO in the ZM-VFN scheme.

\boldmath
\subsection{Virtual Corrections}\label{virtual}
\unboldmath

In the ZM-VFN scheme, where $m_b=0$ is put from the beginning, all singularities including the soft- and collinear gluon emission  are regularized by dimensional regularization in $D=4-2\epsilon$ space-time dimensions to become single poles in $\epsilon$, which are subtracted at factorization scale $\mu_F$ and absorbed into the bare FFs according to the modified minimal-subtraction  scheme ($\overline{MS}$). In this scheme, $m_b$ only sets the initial scale $\mu_F^\text{ini}={\cal O}(m_b)$ of the DGLAP evolution.\\
Adopting the on-shell mass-renormalization scheme, the virtual one-loop corrections to the $tbH^+$-vertex includes both IR- and UV-singularities. Therefore,  the contribution of virtual corrections into the differential decay width normalized to the Born width, reads
\begin{eqnarray}
\frac{1}{\hat\Gamma_0}\frac{d\hat\Gamma^{vir}_b}{dx_b}=\frac{1}{2 a^2 m_t^2(1-y)}\overline{|M^{vir}|^2}\delta(1-x_b),
\end{eqnarray}
where, $a^2=G_Fm_t^2 |V_{tb}|^2\cot^2\beta/\sqrt{2}$, the scaled mass $y$ is $y=m_{H^+}^2/m_t^2$ and \textit{$\overline{|M^{vir}|^2}=1/2\sum_{Spin}(M_0^{\dagger} M_{loop}+M_{loop}^{\dagger} M_0)$}. Following Ref.~\cite{Li}, the renormalized amplitude of the virtual corrections  can be written as
\begin{eqnarray}
M_{loop}=\bar{u_b}(\Lambda_{ct}+\Lambda_l)(a+b\gamma_5)u_t,
\end{eqnarray}
 where  $\Lambda_{ct}$ stands for the counter term and $\Lambda_l$ arises from the one-loop vertex correction.  Since we neglect the b quark mass, thus the counter term of the vertex consists of the top-quark mass renormalization and the wave function renormalizations as
\begin{eqnarray}
\Lambda_{ct}=\frac{1}{2}(\delta Z_b+\delta Z_t)-\frac{\delta m_t}{m_t},
\end{eqnarray}
where, the mass renormalization constant of the top quark reads
\begin{eqnarray}\label{mass}
\frac{\delta m_t}{m_t}=\frac{\alpha_s(\mu_R)}{4\pi}C_F(\frac{3}{\epsilon_{UV}}-3\gamma_E+3\ln\frac{4\pi\mu_F^2}{m_t^2}+4),\qquad
\end{eqnarray}
and from  Ref.~\cite{MoosaviNejad:2009zz}, for the wave function renormalization constants we have
\begin{eqnarray}\label{wave}
\delta Z_t &=& -\frac{\alpha_s(\mu_R)}{4\pi}C_F(\frac{1}{\epsilon_{UV}}+\frac{2}{\epsilon_{IR}}-3\gamma_E+3\ln\frac{4\pi\mu_F^2}{m_t^2}+4),\nonumber\\
\delta Z_b &=&-\frac{\alpha_s(\mu_R)}{4\pi}C_F(\frac{1}{\epsilon_{UV}}-\frac{1}{\epsilon_{IR}}).
\end{eqnarray}
In Eqs.~(\ref{mass}) and (\ref{wave}) , $\epsilon_{IR}$ and $\epsilon_{UV}$ represent infra-red(IR) and ultra-violet(UV) singularities which arise from the collinear- and the soft-gluon singularities, respectively. Therefore, the real part of the one-loop vertex corrections $\Lambda_l$ is given by
\begin{eqnarray}
\Lambda_l &=&\frac{\alpha_s}{4\pi}C_F (-\frac{F^2}{2}-\frac{9}{2}F+2\frac{1+2y}{y}\ln(1-y)-2Li_2(y)\nonumber\\
&&+\frac{2F+1}{2\epsilon_{IR}}+\frac{4}{\epsilon_{UV}}-\frac{1}{\epsilon_{IR}^2}-\frac{49}{8}-\frac{\pi^2}{12}),\nonumber
\end{eqnarray}
where, $F=2\ln(1-y)-\ln(4\pi \mu_F^2/m_t^2)+\gamma_E-5/2$, $C_F=(N_c^2-1)/(2N_c)=4/3$ for $N_c=3$ quark colors,  and $Li_2(x)=-\int_0^x(dt/t)\ln(1-t)$ is the Spence function.\\
  All UV-divergences  are canceled after summing all virtual corrections up but the IR-singularities are remaining which are now shown by $\epsilon$. The virtual corrections to the differential decay rate is then given by
\begin{eqnarray}
\frac{1}{\hat\Gamma_0}\frac{d\hat\Gamma^{vir}_b}{dx_b}&=&\frac{\alpha_s(\mu_R)}{2\pi}C_F\delta(1-x_b)\big(-\frac{1}{\epsilon^2}+\frac{F}{\epsilon}-\frac{F^2}{2}+\nonumber\\
&&(\frac{2}{y}-5)\ln(1-y)-2Li_2(y)-\frac{7}{8}-\frac{\pi^2}{12}\big).\nonumber\\
\end{eqnarray}

\boldmath
\subsection{Real Corrections}\label{real}
\unboldmath

As it is explained in Ref.~\cite{reall}, to cancel the IR-singularities of the virtual corrections, one needs to include the real gluon emission, namely, $t\rightarrow H^+bg$.  To calculate the contribution of the real corrections, we start form the definition of decay rate.  As before, to regulate the IR-divergences we work in $D=4-2\epsilon$  dimensions, therefore from the definition of decay rate, one has 
\begin{eqnarray}\label{decaydef}
d\hat\Gamma^{real}=\frac{\mu_F^{2(4-D)}}{2m_t}\overline{|M^{real}|^2}dPS(p_t, p_b, p_g, p_{H^+}),
\end{eqnarray}
where, the Phase Space element reads
\begin{eqnarray}
dPS&=&\frac{d^{D-1}\bold{p}_b}{(2\pi)^{D-1}2E_b}\frac{d^{D-1}\bold{p}_{H^+}}{(2\pi)^{D-1}2E_{H^+}}\frac{d^{D-1}\bold{p}_g}{(2\pi)^{D-1}2E_g}\nonumber\\
&&\times(2\pi)^D\delta^D(p_t-p_b-p_{H^+}-p_g).
\end{eqnarray}
For simplicity, we choose the top-quark rest-frame and to calculate  the differential rate $d\hat\Gamma^{real}_b/dx_b$ normalized to the Born width, we fix the momentum of b-quark in Eq.~(\ref{decaydef}). To get the correct finite term in the normalized differential decay rate, the Born width ${\Gamma}_0$  will have to be evaluated in the dimensional regularization at ${\cal O}(\epsilon^2)$, i.e.
 $\hat\Gamma_0\rightarrow \hat\Gamma_0\{1-\epsilon(F+1/2)+\epsilon^2(F^2/2+F/2+17/8-\pi^2/4)\}$. Thus, the contribution of the real gluon emission into the normalized differential decay width is then given by
\begin{eqnarray}
\frac{1}{\hat\Gamma_0}\frac{d\hat\Gamma^{real}_b}{dx_b}&=&\frac{\alpha_s}{2\pi}C_F\Big\{\delta(1-x_b)\big[\frac{1}{\epsilon^2}-\frac{1}{\epsilon}(F+\frac{3}{2})+\frac{F^2}{2}+\nonumber\\
&&\frac{3}{2}F-2\frac{y}{1-y}\ln y+2Li_2(1-y)-\frac{\pi^2}{4}+\frac{5}{8}\big]\nonumber\\
&&+\frac{1+x_b^2}{(1-x_b)_+}\big[-\frac{1}{\epsilon}+2\ln x_b+\frac{7x_b^2-8 x_b+7}{2(1+x_b^2)}\nonumber\\
&&+F\big]+2(1+x_b^2)\bigg(\frac{\ln(1-x_b)}{1-x_b}\bigg)_+\Big\},
\end{eqnarray}
where the plus distributions are defined as usual.

\boldmath
\subsection{Analytic Results for Partial Decay Rates}
\unboldmath

The NLO expression for $d\hat\Gamma_b/dx_b$ is obtained by summing the Born term, the virtual one-loop and the real contributions.\\
Since, the B meson can be also produced from the fragmentation of the emitted real gluon, therefore, we also need the differential decay rate $d\hat\Gamma_g/dx_g$ in the ZM-VFN scheme. To calculate the $d\hat\Gamma_g/dx_g$, as before, we start form Eq.~(\ref{decaydef}) by fixing the momentum of gluon, therefore, there will be no soft singularity.\\
Our results list here
\begin{eqnarray}\label{first}
\frac{d\hat\Gamma_b}{dx_b}&=&\hat\Gamma_0\Bigg\{\delta(1-x_b)+\frac{\alpha_s(\mu_R)}{2 \pi}C_F\Big\{\Big[\frac{1+x_b^2}{(1-x_b)_+}+\nonumber\\
&&\frac{3}{2}\delta(1-x_b)\Big]\big(-\frac{1}{\epsilon}+\gamma_E-\ln 4\pi\big)+\hat A_1(x_b)\Big\}\Bigg\},\nonumber\\
\frac{d\hat\Gamma_g}{dx_g}&=&\hat\Gamma_0\frac{\alpha_s(\mu_R)}{2 \pi}C_F\Bigg\{\frac{1+(1-x_g)^2}{x_g}\times\nonumber\\
&&\hspace{2cm}\big(-\frac{1}{\epsilon}+\gamma_E-\ln 4\pi\big)+\hat B_1(x_g)\Bigg\},\nonumber\\
\end{eqnarray}
where  $A_1(x_b)$ and $B_1(x_g)$ are free of singularities and, in the following, their functional form will be shown.\\
To subtract the collinear singularities remaining in Eq.~(\ref{first}), we apply  the modified minimal subtraction ($\overline{MS}$) scheme where the collinear singularities are absorbed into the bare FFs. This renormalizes the FFs and generates the finite terms of the form $\alpha_s\ln(m_t^2/\mu_F^2)$ in the differential decay rates.\\
 According to this scheme, in order to  get the $\overline{MS}$  coefficient functions we shall have to subtract from Eq.~(\ref{first}) the ${\cal O}(\alpha_s)$ term multiplying the  characteristic $\overline{MS}$ constant $(-1/\epsilon+\gamma_E-\ln 4\pi)$, therefore, we obtain
\begin{eqnarray} \label{Final00}
\frac{d\hat\Gamma_b^{\overline{MS}}}{dx_b}&=&\hat\Gamma_0\Big\{\delta(1-x_b)+\nonumber\\
&&\qquad\frac{\alpha_s(\mu_R)}{2 \pi}\Big[P_{qq}^{(0)}(x_b)\ln\frac{m_t^2}{\mu_F^2}+C_F \hat A(x_b)\Big]\Big\},\nonumber\\
\frac{d\hat\Gamma_g^{\overline{MS}}}{dx_g}&=&\hat\Gamma_0\Big\{\frac{\alpha_s(\mu_R)}{2 \pi}\Big[P_{gq}^{(0)}(x_g)\ln\frac{m_t^2}{\mu_F^2}+C_F \hat B(x_g)\Big]\Big\},\nonumber\\
\end{eqnarray}
where  $P_{ab}^{(0)}$ are the time-like splitting functions at leading order \cite{dglap}
\begin{eqnarray}\label{eq:zmvfn}
P_{qq}^{(0)}(z) &=& C_F \bigg(\frac{1+z^2}{(1-z)_+}+\frac{3}{2}\delta(1-z)\bigg),\nonumber\\
P_{gq}^{(0)}(z) &=& C_F \bigg(\frac{1+(1-z)^2}{z}\bigg),\nonumber
\end{eqnarray}
and the coefficient functions $ \hat A(x_b)$ and $\hat B(x_g)$  read
\begin{eqnarray}\label{fin}
\hat A(x_b)&=&\delta(1-x_b)\big[2\frac{1-y}{y}\ln(1-y)-2\frac{y}{1-y}\ln y\nonumber\\
&&-2 Li_2(y)+2 Li_2(1-y)-4-\frac{\pi^2}{3}\big]\nonumber\\
&&+\frac{1+x_b^2}{(1-x_b)_+}\big[\ln(x_b^2(1-y)^2)+\frac{x_b^2-4x_b+1}{1+x_b^2}\big]\nonumber\\
&&+2(1+x_b^2)\bigg(\frac{\ln(1-x_b)}{1-x_b}\bigg)_+,\nonumber\\
\hat B(x_g)&=&\frac{1+(1-x_g)^2}{x_g}\big[\ln(x_g^2(1-y)^2(1-x_g)^2)-\frac{5}{2}\nonumber\\
&&-\ln(1-(1-y)x_g)\big]+\frac{1}{2(1-(1-y)x_g)^2}\big[\nonumber\\
&&7x_g+\frac{6}{x_g}-6-(1-y)(10x_g^2-8x_g+12)+\nonumber\\
&&x_g(1-y)^2(4x_g^2-4x_g+7)\big].\nonumber\\
\end{eqnarray}
In this work we identify $\mu_R=\mu_F=m_t$, so that in Eq.~(\ref{Final00}) the terms proportional to $\ln(m_t^2/\mu_F^2)$ vanish.\\
 Integrating $d\hat\Gamma_b/dx_b$ of Eq.~(\ref{Final00}) over $x_b(0<x_b<1)$, we obtain the renormalized decay rate
\begin{eqnarray}\label{rate}
\hat\Gamma&=&\hat\Gamma_0\Big\{1-\frac{C_F\alpha_s}{2\pi}\bigg[\frac{2y}{1-y}\ln y+(5-\frac{2}{y})\ln(1-y)+\nonumber\\
&&2Li_2(y)-2Li_2(1-y)-\frac{9}{2}+\pi^2\bigg]\Big\}.\nonumber\\
\end{eqnarray}
 This result is in agreement with Refs.~\cite{Czarnecki:1992ig,Liu:1992qd} and the corrected version of \cite{Li}. As it is seen from Eq.~(\ref{rate}), in the limit $m_{H^+}/m_t\rightarrow 0(\equiv y\rightarrow 0)$ the total decay rate is finite and proportional to $G_Fm_t^3$. At the opposite limit where  $m_{H^+}/m_t\rightarrow 1(\equiv y\rightarrow 1)$, due to the $\ln(1-y)$ singularity   setting the bottom-quark mass to be zero is no longer a valid approximation for the differential decay rates and our results must be improved considering a massive b-quark.

\section{Numerical analysis}
\label{sec:four}

We are now in a position to present our phenomenological results by performing a numerical analysis. In the  MSSM, the mass of the charged Higgs is strongly correlated with the other Higgs boson masses. The charged Higgs boson mass in the MSSM is restricted at tree-level by $m_{H^+}>m_W$, Ref.~\cite{Nakamura:2010zzi}. This restriction does not hold for some regions of parameter space after including radiative corrections. Therefore, two key phenomenological parameters in the charged Higgs searches are the  Higgs mass $m_{H^+}$ and $\tan\beta$, which are model-dependent. Searches of the charged Higsses over a good part of the $m_{H^+}-\tan\beta$ plane in the MSSM is a program that still has to be carried out and this belongs to the LHC experiments. We adopt from Ref.~\cite{Nakamura:2010zzi} the present  limit $m_{H^+}>79.3$ GeV obtained from LEP. We also adopt from Ref.~\cite{Nakamura:2010zzi} the input parameter values
$G_F = 1.16637\times10^{-5}$~GeV$^{-2}$,
$m_t = 172.0$~GeV,
$m_b = 4.90$~GeV, and
$m_B = 5.279$~GeV.
We evaluate $\alpha_s^{(n_f)}(\mu_R)$ at NLO in the $\overline{\text{MS}}$
scheme using Eq.~(8) of Ref.~\cite{Corcella:1}, with $n_f=5$ active quark flavors and
the typical QCD scale $\Lambda_{\overline{\text{MS}}}^{(5)}=231.0$~MeV. As mentioned before,  the b-quark mass only enter through the initial condition of the nonperturbative fragmentation function.
We employ the nonperturbative $B$-hadron  FFs that were determined at NLO in
the ZM-VFN scheme through a joint fit \cite{Kniehl:2008zza} to
$e^+e^-$-annihilation data taken by ALEPH \cite{Heister:2001jg} and OPAL
\cite{Abbiendi:2002vt} at CERN LEP1 and by SLD \cite{Abe:1999ki} at SLAC SLC.
Specifically, the power ansatz $D_b(z,\mu_F^\text{ini})=Nz^\alpha(1-z)^\beta$
was used as the initial condition for the $b\to B$ FF at
$\mu_F^\text{ini}=4.5$~GeV, while the gluon and light-quark FFs were generated
via the DGLAP evolution.
The fit yielded $N=4684.1$, $\alpha=16.87$, and $\beta=2.628$.

\begin{figure}
\begin{center}
\includegraphics[width=1\linewidth,bb=37 192 552 629]{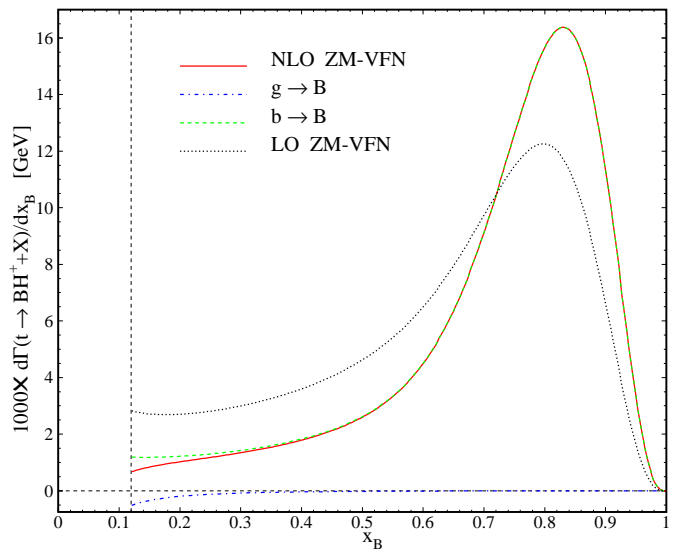}
\caption{\label{fig1}%
$d\Gamma(t\to BH^++X)/dx_B$ as a function of $x_B$ in the ZM-VFN ($m_b=0$)
scheme with $m_{H^+}=120$  GeV and $\tan\beta=10$.
The NLO result (solid line) is compared to the LO one (dotted line) and broken
up into the contributions due to $b\to B$ (dashed line) and $g\to B$
(dot-dashed line) fragmentation.}
\end{center}
\end{figure}

To study the scaled-energy ($x_B$) distribution of the bottom-flavored hadrons produced in top-quark decay, we consider the quantity $d\Gamma(t\to BH^++X)/dx_B$. In Fig.~\ref{fig1}, we show our prediction for the size of the NLO corrections, by comparing the LO (dotted line) and NLO (solid line) results,
and the relative importance of the $b\to B$ (dashed line) and $g\to B$
(dot-dashed line) fragmentation channels at NLO, taking $\tan\beta=10$ and $m_{H^+}=120$ GeV. The same NLO FFs are used for the LO result.
Fig.~\ref{fig1} shows that the NLO corrections lead to a
significant enhancement of the decay rate in the peak region and above.
Furthermore, the peak position is shifted towards higher values of $x_B$. The gluon fragmentation leads to an  appreciable reduction in decay rate at low-$x_B$ region, for $x_B\alt0.3$.  For example, the gluon fragmentation decreases the size of decay rate up to $43\%$ at  $x_B=0.12$.  
For higher values of $x_B$,  the $b\to B$ contribution is dominant. As we explained in section~\ref{sec:one}, the mass of B-meson is responsible for the appearance of the threshold at $x_B=2m_B/(m_t(1-y))=0.12$.

\begin{figure}
\begin{center}
\includegraphics[width=1\linewidth,bb=37 192 552 629]{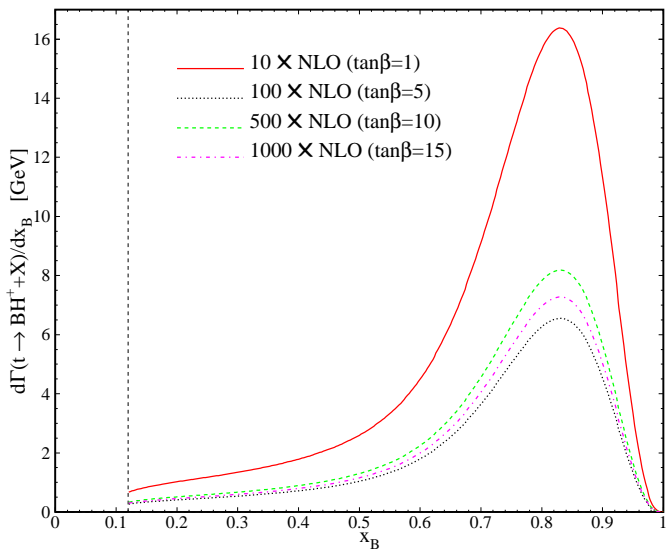}
\caption{\label{fig2}%
$d\Gamma(t\to BH^++X)/dx_B$ as a function of $x_B$ in different values of $\tan\beta=1,5,10$ and $15$, with $m_{H^+}=120$ GeV. When the values of $\tan\beta$ exceed $\tan\beta=2$, the decay rate becomes quite small \cite{kadeer}.}
\end{center}
\end{figure}

In Fig.~(2b) of Ref.~\cite{kadeer}, the unpolarized rate is plotted as a function of $\tan\beta$ for a sample value of $m_{H^+}=120$ GeV. It can be seen that the rate is quite small for $\tan\beta$ values exceeding $\tan\beta=2$.\\
  In Fig.~\ref{fig2}, we study the energy distribution of the B-meson in  different values of the $\tan\beta$, i.e. $\tan\beta=1, 5, 10$ and 15, for $m_{H^+}=120$ GeV. It can be seen that when $\tan\beta$ is increased the decay rate is decreased, as $\hat\Gamma_0$ is proportional to $\cot^2\beta$.

In Fig.~(2a) of Ref.~\cite{kadeer}, the unpolarized rate is also shown  as a function of $y=m_{H^+}/m_t$ for  $\tan\beta=10$. The  functional behavior of the rate shows that the rate is largest  when  $m_{H^+}\rightarrow 0$ and it drops to zero when $m_{H^+}\rightarrow  m_t$.

\begin{figure}
\begin{center}
\includegraphics[width=1\linewidth,bb=37 192 552 629]{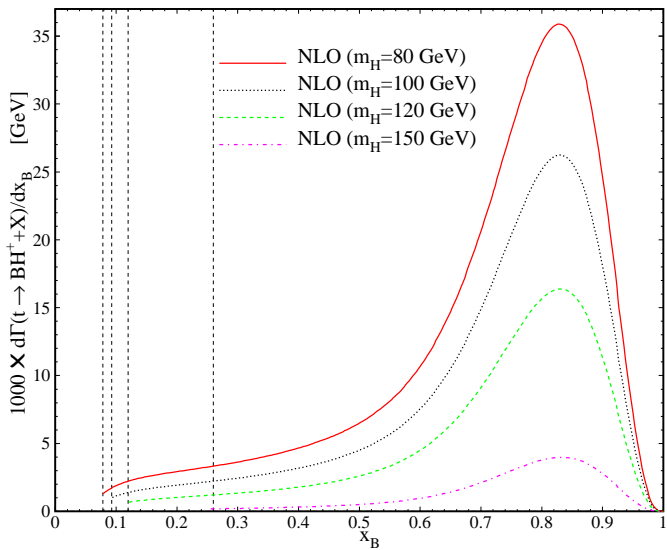}
\caption{\label{fig3}%
$d\Gamma(t\to BH^++X)/dx_B$ as a function of $x_B$  in the ZM-VFN
scheme with different values of $m_{H^+}=80, 100, 120$ and 150 GeV ($\tan\beta=10$). Thresholds at $x_B$ are shown.}
\end{center}
\end{figure}

\begin{figure}
\begin{center}
\includegraphics[width=1\linewidth,bb=37 192 552 629]{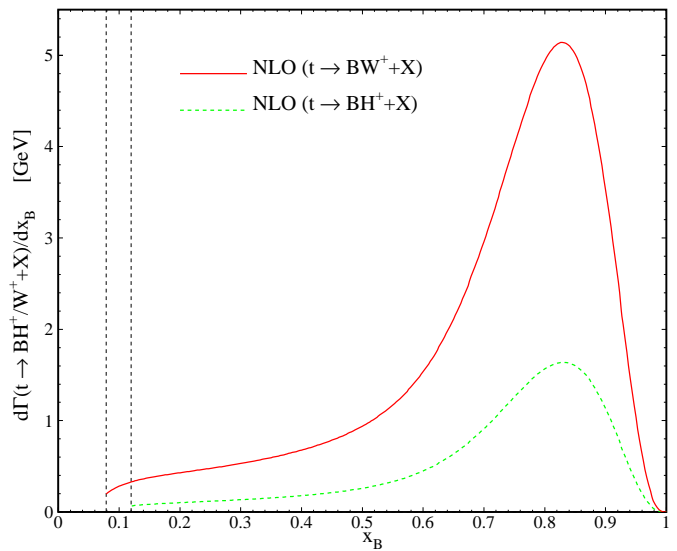}
\caption{\label{fig4}%
$x_B$ spectrum in top decay considering the decay modes  $t\rightarrow BW^++X$ (solid line) and $t\rightarrow BH^++X$  (dashed line), taking $m_{W^+}=80.399$ GeV, $m_{H^+}=120$  GeV and $\tan\beta=1$. }
\end{center}
\end{figure}

 Adopting the limit $m_{H^+}>79.3$ GeV from Ref.~\cite{Nakamura:2010zzi}, in Fig.~\ref{fig3}  we study the energy distribution of the B-meson in  different values of the Higgs boson mass, i.e. $m_{H^+}=80, 100, 120$ and 150 GeV, by fixing  $\tan\beta=10$. As mentioned,  the mass of B-meson creates the thresholds at $x_B=0.08$ (for $m_{H^+}=80$ GeV), $x_B=0.09$ (for $m_{H^+}=100$ GeV), $x_B=0.12$ (for $m_{H^+}=120$ GeV) and $x_B=0.26$ (for $m_{H^+}=150$ GeV).

In comparison with the  Born rate $\Gamma_{t\rightarrow b+W^+}=1.364$ GeV,  the rate into a charged Higgs is generally quite small except for small $\tan\beta$ values. One finds equality of the rates into a $W^+$ and $H^+$ only at $\tan\beta=0.56$ for $m_{H^+}=120$ GeV. However, such a small $\tan\beta$ value is excluded by the indirect limits in the $(m_{H^\pm}, \tan\beta)$ plane \cite{pdg}.

In Fig.~\ref{fig4}, we compared the energy distribution of the B-meson produced in decay modes $t\rightarrow BW^++X$ and $t\rightarrow BH^++X$ with  $\tan\beta=1$, $m_{H^+}=120$ GeV,  $m_W=80.399$ GeV and  the parton-level differential  rates $d\hat\Gamma_b(t\rightarrow bW^+)/dx_b$ and  $d\hat\Gamma_g(t\rightarrow bW^+)/dx_g$ extracted from Ref.~\cite{MoosaviNejad:2009zz}. The thresholds appear at $x_B=0.08$ (for $t\rightarrow BW^++X$) and $x_B=0.12$ ( for $t\rightarrow BH^++X$).\\
 The total top-quark decay width is obtained by summing the two partial widths $\Gamma_{t\rightarrow b+W^+}$ and $\Gamma_{t\rightarrow b+H^+}$ order by order. However, as Figs.~(\ref{fig2}) and (\ref{fig3}) show  the radiative corrections coming from the supersymmetric  sector  depend extremely on the Higgs mass and $\tan\beta$, but Fig.~\ref{fig4} shows that the contribution of the decay mode $t\rightarrow W^++b$  is always larger than the one coming from  $t\rightarrow H^++b$, see more detail in Ref.~\cite{Ali:2011qf}.

\section{Conclusions}
\label{sec:five}

Top-quark decays within the Standard Model are completely dominated by the mode $t\rightarrow W^++b$ due to $|V_{tb}|=1$ to a very high accuracy. In beyond-the-SM theories with an extended Higgs sector, top-quark decays can also be  done  via $t\rightarrow H^++b$. This charged Higgs boson has been searched for in high energy experiments, in particular, at LEP and the Tevatron but it has not been seen so far. To study the Higgs boson and new physics beyond the SM at LHC, as a superlative top factory, we need to understand the energy distribution of B-meson produced from top-quark decay. The dominant decay mode  $t\rightarrow BW^++X$ was studied in Refs.~\cite{Corcella:1} and \cite{MoosaviNejad:2009zz}. In this paper we studied the scaled-energy ($x_B$) distribution of B-meson in  $t\rightarrow BH^++X$ at NLO in the ZM-VFN scheme. We employed the nonperturbative B-meson FFs determined by a global fit \cite{Kniehl:2008zza} of experimental data from Z factories, relying on their universality and scaling violations \cite{collins}.

Comparison of  future measurements of $d\Gamma/dx_B$ at the LHC with our NLO predictions   will be important for future tests of the Higgs coupling in the minimal supersymmetric SM(MSSM) and it will be the primary source of information on the B-meson FFs.

\begin{acknowledgments}
I would like to thank Professor Gustav Kramer for reading and improving the manuscript and also for important discussions and comments.  I  would also like to thank Professor Bernd A. Kniehl for his helpful advices and oppening the doors of perturbative QCD to me. This work was supported by Yazd university and the Institute for Research in Fundamental
Science (IPM).
\end{acknowledgments}


\begin{thebibliography}{25}
\bibitem{Nakamura:2010zzi}
  K.~Nakamura {\it et al.}\  (Particle Data Group),
  %``Review of particle physics,''
  J.\ Phys.\ G {\bf 37}, 075021 (2010).
  %%CITATION = JPHGB,G37,075021;%%
  %\cite{Ali:2011qf}

\bibitem{Cabibbo:1963yz}
  N.~Cabibbo,
  %``Unitary Symmetry and Leptonic Decays,''
  Phys.\ Rev.\ Lett.\  {\bf 10}, 531 (1963);
%  %%CITATION = PRLTA,10,531;%%
%\cite{Kobayashi:1973fv}
%\bibitem{Kobayashi:1973fv}
  M.~Kobayashi and T.~Maskawa,
  %``CP Violation In The Renormalizable Theory Of Weak Interaction,''
  Prog.\ Theor.\ Phys.\  {\bf 49}, 652 (1973).
  %%CITATION = PTPKA,49,652;%%

%\cite{Bernreuther:2008ju}
%\cite{Bernreuther:2008ju}
\bibitem{Bernreuther:2008ju}
  W.~Bernreuther,
  %``Top quark physics at the LHC,''
  J.\ Phys.\ G {\bf 35}, 083001 (2008).
%  [arXiv:0805.1333 [hep-ph]].
  %%CITATION = JPHGB,G35,083001;%%

%\cite{Jezabek:1988iv}

%\cite{Gunion:1984yn}
\bibitem{Gunion}
  J.~F.~Gunion and H.~E.~Haber,
  %``Higgs Bosons In Supersymmetric Models. 1,''
  Nucl.\ Phys.\  B {\bf 272} (1986) 1
  [Erratum-ibid.\  B {\bf 402} (1993) 567].
  %%CITATION = NUPHA,B272,1;%%


\bibitem{Li}
  C.~S.~Li and T.~C.~Yuan,
  %``QCD Correction To Charged Higgs Decay Of The Top Quark,''
  Phys.\ Rev.\  D {\bf 42}, 3088 (1990)
  [Erratum-ibid.\  D {\bf 47}, 2156 (1993\ PHRVA,D47,2156.1993)].
  %%CITATION = PHRVA,D47,2156;%%


\bibitem{ATLAS}
ATLAS Collaboration, \textit{Search for charged Higgs bosons in the $\tau+$jets final state in $t\bar t$ decays with $1.03 fb^{-1}$ of $pp$ collision data recorded at $\sqrt{s}=7$ TeV with the ATLAS experiment}, ATLAS-CONF-2011-138 (2011)


\bibitem{Ali:2011qf}
  A.~Ali, F.~Barreiro and J.~Llorente,
  %``Improved sensitivity to charged Higgs searches in Top quark decays $t \to
  %bH^+ \to b (\tau^+\nu_\tau)$ at the LHC using $\tau$ polarisation and
  %multivariate tecnniques,''
  arXiv:1103.1827 [hep-ph].
  %%CITATION = ARXIV:1103.1827;%%

\bibitem{Czarnecki:1998qc}
  A.~Czarnecki and K.~Melnikov,
  %``Two-loop {QCD} corrections to top quark width,''
  Nucl.\ Phys.\  {\bf B544}, 520 (1999);
%  [arXiv:hep-ph/9806244].
  %%CITATION = NUPHA,B544,520;%%
%\cite{Chetyrkin:1999ju}
%\bibitem{Chetyrkin:1999ju}
  K.~G.~Chetyrkin, R.~Harlander, T.~Seidensticker, and M.~Steinhauser,
  %``Second order {QCD} corrections to Gamma(t --> W b),''
  Phys.\ Rev.\  D {\bf 60}, 114015 (1999);
%  [arXiv:hep-ph/9906273].
  %%CITATION = PHRVA,D60,114015;%%
%\cite{Blokland:2004ye}
%\bibitem{Blokland:2004ye}
  I.~R.~Blokland, A.~Czarnecki, M.~\'Slusarczyk, and F.~Tkachov,
  %``Heavy-to-light decays with a two-loop accuracy,''
  Phys.\ Rev.\ Lett.\  {\bf 93}, 062001 (2004);
%  [arXiv:hep-ph/0403221].
  %%CITATION = PRLTA,93,062001;%%
%\cite{Blokland:2005vq}
%\bibitem{Blokland:2005vq}
%  I.~R.~Blokland, A.~Czarnecki, M.~Slusarczyk and F.~Tkachov,
  %``Next-to-next-to-leading order calculations for heavy-to-light decays,''
  Phys.\ Rev.\  D {\bf 71}, 054004 (2005);
%  [Erratum-ibid.\  D
{\bf 79}, 019901(E) (2009);
%]
%  [arXiv:hep-ph/0503039].
  %%CITATION = PHRVA,D71,054004;%%
%\cite{Bonciani:2008wf}
%\bibitem{Bonciani:2008wf}
  R.~Bonciani and A.~Ferroglia,
  %``Two-Loop QCD Corrections to the Heavy-to-Light Quark Decay,''
  JHEP {\bf 0811}, 065 (2008).
%  [arXiv:0809.4687 [hep-ph]].
  %%CITATION = JHEPA,0811,065;%%

%\cite{Mehen:1997mw}
\bibitem{Denner:1990ns}
  A.~Denner and T.~Sack,
  %``The Top width,''
  Nucl.\ Phys.\  B {\bf 358}, 46 (1991);
  %%CITATION = NUPHA,B358,46;%%
%\cite{Eilam:1991iz}
%\bibitem{Eilam:1991iz}
  G.~Eilam, R.~R.~Mendel, R.~Migneron, and A.~Soni,
  %``Radiative corrections to top quark decay,''
  Phys.\ Rev.\ Lett.\  {\bf 66}, 3105 (1991);
  %%CITATION = PRLTA,66,3105;%%
%\cite{Yuan:1991av}
%\bibitem{Yuan:1991av}
  C.-P.~Yuan and T.~C.~Yuan,
  %``Leading electroweak radiative corrections to $t \to W^{+}$ + $b$,''
  Phys.\ Rev.\  D {\bf 44}, 3603 (1991);
  %%CITATION = PHRVA,D44,3603;%%
%\cite{Kuruma:1992if}
%\bibitem{Kuruma:1992if}
  T.~Kuruma,
  %``Electroweak radiative corrections to the top quark decay,''
  Z.\ Phys.\  C {\bf 57}, 551 (1993);
  %%CITATION = ZEPYA,C57,551;%%
%\cite{Oliveira:2001vw}
%\bibitem{Oliveira:2001vw}
  S.~M.~Oliveira, L.~Br\"ucher, R.~Santos, and A.~Barroso,
  %``Electroweak corrections to the top quark decay,''
  Phys.\ Rev.\  D {\bf 64}, 017301 (2001).
%  [arXiv:hep-ph/0011324].
  %%CITATION = PHRVA,D64,017301;%%

%\cite{Corcella:2001hz}

\bibitem{Corcella:1}
  G.~Corcella and A.~D.~Mitov,
  %``Bottom quark fragmentation in top quark decay,''
  Nucl.\ Phys.\  B {\bf 623}, 247 (2002).
%  [arXiv:hep-ph/0110319].
  %%CITATION = NUPHA,B623,247;%%

%\cite{Corcella:2009rs}
\bibitem{Corcella:2}
  G.~Corcella and F.~Mescia,
  %``A Phenomenological Study of Bottom Quark Fragmentation in Top Quark
  %Decay,''
  Eur.\ Phys.\ J.\  C {\bf 65}, 171 (2010);
%  [Erratum-ibid.\  C
{\bf 68}, 687(E) (2010);
%]
%  [arXiv:0907.5158 [hep-ph]].
  %%CITATION = EPHJA,C65,171;%%
%\bibitem{Biswas:2010sa}
  S.~Biswas, K.~Melnikov, and M.~Schulze,
  %``Next-to-leading order QCD effects and the top quark mass measurements at
  %the LHC,''
  JHEP {\bf 1008}, 048 (2010).
%  [arXiv:1006.0910 [hep-ph]].
  %%CITATION = JHEPA,1008,048;%%

%\cite{Mele:1990cw}
\bibitem{do}
D$\varnothing$  Collaboration,~B.~Abbott et al.,
Phys.\ Rev.\  D {\bf 58}, 052001 (1998);\\
 CDF Collaboration,~T.~Affolder et al.,
 Phys.\ Rev.\  D {\bf 63}, 032003 (2001).

\bibitem{jm}
J.~Binnewies,~B.A.~Kniehl,~and G.~Kramer,
Phys.\ Rev.\  D {\bf 58}, 034016 (1998);\\
 M.~Cacciari and M.~Greco,~Nucl.~\ Phys.~B{\bf 421},~530(1994).

\bibitem{dglap}
  V.~N.~Gribov and L.~N.~Lipatov,
  %``Deep Inelastic E P Scattering In Perturbation Theory,''
  Sov.\ J.\ Nucl.\ Phys.\  {\bf 15}, 438 (1972)
  [Yad.\ Fiz.\  {\bf 15}, 781 (1972)];
  %%CITATION = YAFIA,15,781;%%
%\cite{Altarelli:1977zs}
%\bibitem{Altarelli:1977zs}
  G.~Altarelli and G.~Parisi,
  %``Asymptotic Freedom In Parton Language,''
  Nucl.\ Phys.\ {\bf B126}, 298 (1977);
  %%CITATION = NUPHA,B126,298;%%
%\cite{Dokshitzer:1977sg}
%\bibitem{Dokshitzer:1977sg}
  Yu.~L.~Dokshitzer,
  %``Calculation Of The Structure Functions For Deep Inelastic Scattering And E+
  %E- Annihilation By Perturbation Theory In Quantum Chromodynamics,''
  Sov.\ Phys.\ JETP {\bf 46}, 641 (1977)
  [Zh.\ Eksp.\ Teor.\ Fiz.\  {\bf 73}, 1216 (1977)].
  %%CITATION = ZETFA,73,1216;%%

%\cite{Kneesch:2007ey}

\bibitem{jc}
J.~C.~Collins,~Phys. \ Rev.\ D~{\bf 66} (1998) 094002.

\bibitem{higg}
J.~F.~Gunion, H.~Haber, G.~Kane, and S.~Dawson,\textit{ The Higgs Hunter's Guide} (Addison-Wesley, Reading, MAA, 1990), and refrences therein.

\bibitem{kadeer}
  A.~Kadeer, J.~G.~K\"orner, and M.~C.~Mauser,
  %``A Phenomenological Study of Bottom Quark Fragmentation in Top Quark
  %Decay,''
  Eur.\ Phys.\ J.\  C {\bf 54}, 175 (2008).
  %\cite{Li:1990cp}
\bibitem{MoosaviNejad:2009zz}
  S.~M.~Moosavi Nejad,
 ``Bottom-hadron production through top quark decay,''
 %CITATION =
  DESY-THESIS-2009-017;%%

\bibitem{reall}
  T.~Kinosita, J.~Math.~Phys.~3,~650 (1962); T.~D.~Lee and M.~ Nauenberg,~Phys.\ Rev.\ B~{\bf 1549} (1964) 133.
  %\cite{Li:1990cp}

%\cite{Czarnecki:1992ig}
%\cite{MoosaviNejad:2009zz}
%\cite{MoosaviNejad:2009zz}

\bibitem{Czarnecki:1992ig}
  A.~Czarnecki and S.~Davidson,
  %``On the QCD corrections to the charged Higgs decay of a heavy quark,''
  Phys.\ Rev.\  D {\bf 47}, 3063 (1993)
  [arXiv:hep-ph/9208240].
  %%CITATION = PHRVA,D47,3063;%%
  %\cite{Liu:1992qd}
\bibitem{Liu:1992qd}
  J.~Liu and Y.~P.~Yao,
  %``QCD corrections to the charged Higgs boson decay of a heavy top quark,''
  Phys.\ Rev.\  D {\bf 46}, 5196 (1992)
  [arXiv:hep-ph/9205245].
  %%CITATION = PHRVA,D46,5196;%%

  %\cite{Nakamura:2010zzi}
\bibitem{Kniehl:2008zza}
  B.~A.~Kniehl, G.~Kramer, I.~Schienbein, and H.~Spiesberger,
  %``Finite-mass effects on inclusive $B$ meson hadroproduction,''
  Phys.\ Rev.\  D {\bf 77}, 014011 (2008).
%  [arXiv:0705.4392 [hep-ph]].
  %%CITATION = PHRVA,D77,014011;%%


%\cite{Heister:2001jg}
\bibitem{Heister:2001jg}
  A.~Heister {\it et al.}\  (ALEPH Collaboration),
  %``Study of the fragmentation of b quarks into B mesons at the Z peak,''
  Phys.\ Lett.\  B {\bf 512}, 30 (2001).
%  [arXiv:hep-ex/0106051].
  %%CITATION = PHLTA,B512,30;%%

%\cite{Abbiendi:2002vt}
\bibitem{Abbiendi:2002vt}
  G.~Abbiendi {\it et al.}\  (OPAL Collaboration),
  %``Inclusive analysis of the b quark fragmentation function in Z decays at
  %LEP. ((B)),''
  Eur.\ Phys.\ J.\  C {\bf 29}, 463 (2003).
%  [arXiv:hep-ex/0210031].
  %%CITATION = EPHJA,C29,463;%%

%\cite{Abe:1999ki}
\bibitem{Abe:1999ki}
  K.~Abe {\it et al.}\  (SLD Collaboration),
  %``Precise measurement of the b-quark fragmentation function in Z0 boson
  %decays,''
  Phys.\ Rev.\ Lett.\  {\bf 84}, 4300 (2000);
%  [arXiv:hep-ex/9912058].
  %%CITATION = PRLTA,84,4300;%%
%\cite{Abe:2002iq}
%\bibitem{Abe:2002iq}
%  K.~Abe {\it et al.}  [SLD Collaboration],
  %``Measurement of the b-quark fragmentation function in Z0 decays,''
  Phys.\ Rev.\  D {\bf 65}, 092006 (2002);
%  [Erratum-ibid.\  D
{\bf 66}, 079905(E) (2002).
%]
%  [arXiv:hep-ex/0202031].
  %%CITATION = PHRVA,D65,092006;%%

\bibitem{pdg}
Particle Data Group, W.~M.~ Yao et al, J.\ Phys.\ G {\bf 33}, (2006) 1.

\bibitem{collins}
  J.~C.~Collins,
  %``Finite-mass effects on inclusive $B$ meson hadroproduction,''
  Phys.\ Rev.\  D {\bf 58}, 094002 (1998).



\end{thebibliography}
\end{document}